\documentclass[graybox]{svmult}

\usepackage{mathptmx}       
\usepackage{helvet}         
\usepackage{courier}        
\usepackage{type1cm}        
\usepackage{makeidx}         
\usepackage{graphicx}        
\usepackage{multicol}        
\usepackage[bottom]{footmisc}

\makeindex             

\begin{document}

\title*{Finding Community Structure Based on Subgraph Similarity}
\author{Biao Xiang, En-Hong Chen, and Tao Zhou}
\institute{Biao Xiang, En-Hong Chen \at Department of Computer
Science, University of Science and Technology of China, Hefei Anhui
230009, P. R. China. \email{cheneh@ustc.edu.cn} \and Tao Zhou \at
Department of Modern Physics, University of Science and Technology
of China, Hefei Anhui 230026, P. R. China, and Department of
Physics, University of Fribourg, Chemin du Mus\'ee 3, Fribourg 1700,
Switzerland. \email{zhutou@ustc.edu}}
%
%
\maketitle

\abstract{Community identification is a long-standing challenge in
the modern network science, especially for very large scale networks
containing millions of nodes. In this paper, we propose a new metric
to quantify the structural similarity between subgraphs, based on
which an algorithm for community identification is designed.
Extensive empirical results on several real networks from disparate
fields has demonstrated that the present algorithm can provide the
same level of reliability, measure by modularity, while takes much
shorter time than the well-known fast algorithm proposed by Clauset,
Newman and Moore (CNM). We further propose a hybrid algorithm that
can simultaneously enhance modularity and save computational time
compared with the CNM algorithm.}

\section{Introduction}

The study of complex networks has become a common focus of many
branches of science \cite{Newman2006}. An open problem that attracts
increasing attention is the identification and analysis of
communities \cite{Newman2006b}. The so-called communities can be
loosely defined as distinct subsets of nodes within which they are
densely connected, while sparser between which \cite{Girvan2002}.
The knowledge of community structure is significant for the
understanding of network evolution \cite{Palla2007} and the dynamics
taking place on networks, such as epidemic spreading
\cite{Liu2005,Yan2007} and synchronization
\cite{Arenas2006,Zhou2007}. In addition, reasonable identification
of communities is helpful for enhancing the accuracy of information
filtering and recommendation \cite{Xue2005}.

Many algorithms for community identification have been proposed,
these include the agglomerative method based on node similarity
\cite{Breiger1975}, divisive method via iterative removal of the
edge with the highest betweenness \cite{Girvan2002,Newman2004},
divisive method based on dissimilarity index between
nearest-neighboring nodes \cite{Zhou2003}, a local algorithm based
on edge-clustering coefficient \cite{Radicchi2004}, Potts model for
fuzzy community detection \cite{Reichardt2004}, simulated annealing
\cite{Guimera2004}, extremal optimization \cite{Duch2005},
spectrum-based algorithm \cite{Newman2006c}, iterative algorithm
based on passing message \cite{Frey2007}, and so on.

Finding out the optimal division of communities, measure by
modularity \cite{Newman2004}, is very hard \cite{Brandes2007}, and
for most cases, we can only get the near optimal division. Generally
speaking, without any prior knowledge, such as the maximal community
size and the number of communities, an algorithm that can give
higher modularity is more time consuming \cite{Danon2005}. As a
consequence, providing accurate division of communities for a very
large scale network in reasonable time is a big challenge in the
modern network science. To address this issue, Newman proposed a
fast greedy algorithm with time complexity $O(n^2)$ for sparse
networks \cite{Newman2004b}, where $n$ denotes the number of nodes.
Furthermore, Clauset, Newman, and Moore (CNM) designed an improved
algorithm giving identical result but with lower computational
complexity \cite{Clauset2004}, as $O(n\texttt{log}^2n)$. In this
paper, based on a newly proposed metric of similarity between
subgraphs, we design an agglomerative algorithm for community
identification, which gives the same level of reliability but is
typically hundreds of times faster than the CNM algorithm. We
further propose a hybrid method that can simultaneously enhance
modularity and save computational time compared with the CNM
algorithm.

The rest of this paper is organized as follows. In Section 2, we
introduce the present method, including the new metric of subgraph
similarity and the corresponding algorithm, as well as the hybrid
algorithm. In Section 3, we give a brief description of the
empirical data used in this paper. The performance of our proposed
algorithms for both algorithmic accuracy and computational time are
presented in Section 4. Finally, we sum up this paper in Section 5.

\section{Method}

Considering an undirected simple network $G(V,E)$, where $V$ is the
set of nodes and $E$ is the set of edges. The multiple edges and
self-connections are not allowed. Denote
$\Gamma=\{V_1,V_2,\cdots,V_h\}$ a division of $G$, that is, $V_i\cap
V_j=\emptyset$ for $1\leq i \neq j \leq h$ and $V_1\cup V_2 \cup
\cdots \cup V_h=V$. We here propose a new metric of similarity
between two subgraphs, $V_i$ and $V_j$, as:
\begin{equation}
s_{ij}=\frac{e_{ij}+\sum^h_{k=1}\frac{\sqrt{e_{ik}e_{kj}}}{|V_k|}}{\sqrt{d_id_j}},
\end{equation}
where $e_{ij}$ is the number of edges with two endpoints
respectively belonging to $V_i$ and $V_j$ ($e_{ij}$ is defined to be
zero if $i=j$), $|V_k|$ is the number of nodes in subgraph $V_k$,
and $d_i=\sum_{x\in V_i}k_x$ is the sum of degrees of nodes in
$V_i$, where the degree of node $x$, namely $k_x$, is defined as the
number of edges adjacent to $x$ in $G(V,E)$. The similarity here can
be considered as a measure of proximity between subgraphs, and two
subgraphs having more connections or being simultaneously closely
connected to some other subgraphs are supposed to have higher
proximity to each other. $d_i$ can be considered as the mass of a
subgraph, and the denominator, $\sqrt{d_id_j}$, is introduced to
reduce the bias induced by the inequality of subgraph sizes. Note
that, if each subgraph only contains a single node, as
$V_i=\{v_i\}$, the similarity between too subgraphs, $V_i$ and
$V_j$, is degenerated to the well-known Salton index (also called
cosine similarity in the literature) \cite{Salton1983} between $v_i$
and $v_j$ if they are not directly connected.

Our algorithm starts from an $n$-division
$\Gamma_0=\{V_1,V_2,\cdots,V_n\}$ with $V_i=\{v_i\}$ for $1\leq i
\leq n$. The procedure is as follows. (i) For each subgraph $V_i$,
let it connect to the most similar subgraphs, namely
$\{V_j|s_{ij}=\texttt{max}_k\{s_{ik}\}\}$. (ii) Merge each connected
component in the network of subgraphs generated by step (i) into one
subgraph, which defines the next division. (iii) Repeat the step (i)
until the number of subgraphs equals one. During this procedure, we
calculate the modularity for each division and the one corresponding
to the maximal modularity is recorded. To make our algorithm clear
to readers, we show a small scale example consisted of six subgraphs
with similarity matrix:
\begin{equation}
S=\left(
\begin{tabular} {cccccc}
0 & 2 & 2 & 1 & 0 & 1 \\
2 & 0 & 1 & 3 & 1 & 1 \\
2 & 1 & 0 & 1 & 0 & 1 \\
1 & 3 & 1 & 0 & 2 & 0 \\
0 & 1 & 0 & 2 & 0 & 3 \\
1 & 1 & 1 & 0 & 3 & 0 \\
\end{tabular}
\right).
\end{equation}
After the step (i), as shown in Figure 1, we get a network where
each node represents a subgraph. We use the directed network
representation, in which a directed arc from $V_i$ to $V_j$ means
$V_j$ is one of the most similar subgraphs to $V_i$. In the
algorithmic implementation, those directed arcs can be treated as
undirected (symmetry) edges. The network shown in Figure 1 is
determined by the similarity matrix $S$, and after step (ii), the
updated division contains only two subgraphs, $V_1\cup V_2 \cup V_3
\cup V_4$ and $V_5\cup V_6$, corresponding to the two connected
components. Note that, the algorithmic procedure is deterministic
and the result is therefore not sensitive to where it starts at all.

\begin{center}
\begin{figure}
\sidecaption
\includegraphics[scale=0.35]{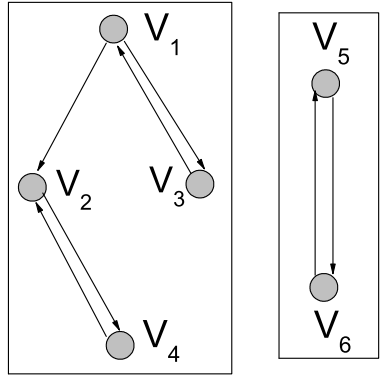}
\caption{Illustration of the algorithm procedure, where each node
represents a subgraph. The similarities between subgraph pairs are
shown in Eq. (2).}
\end{figure}
\end{center}

\begin{center}
\begin{figure}
\sidecaption
\includegraphics[scale=0.30]{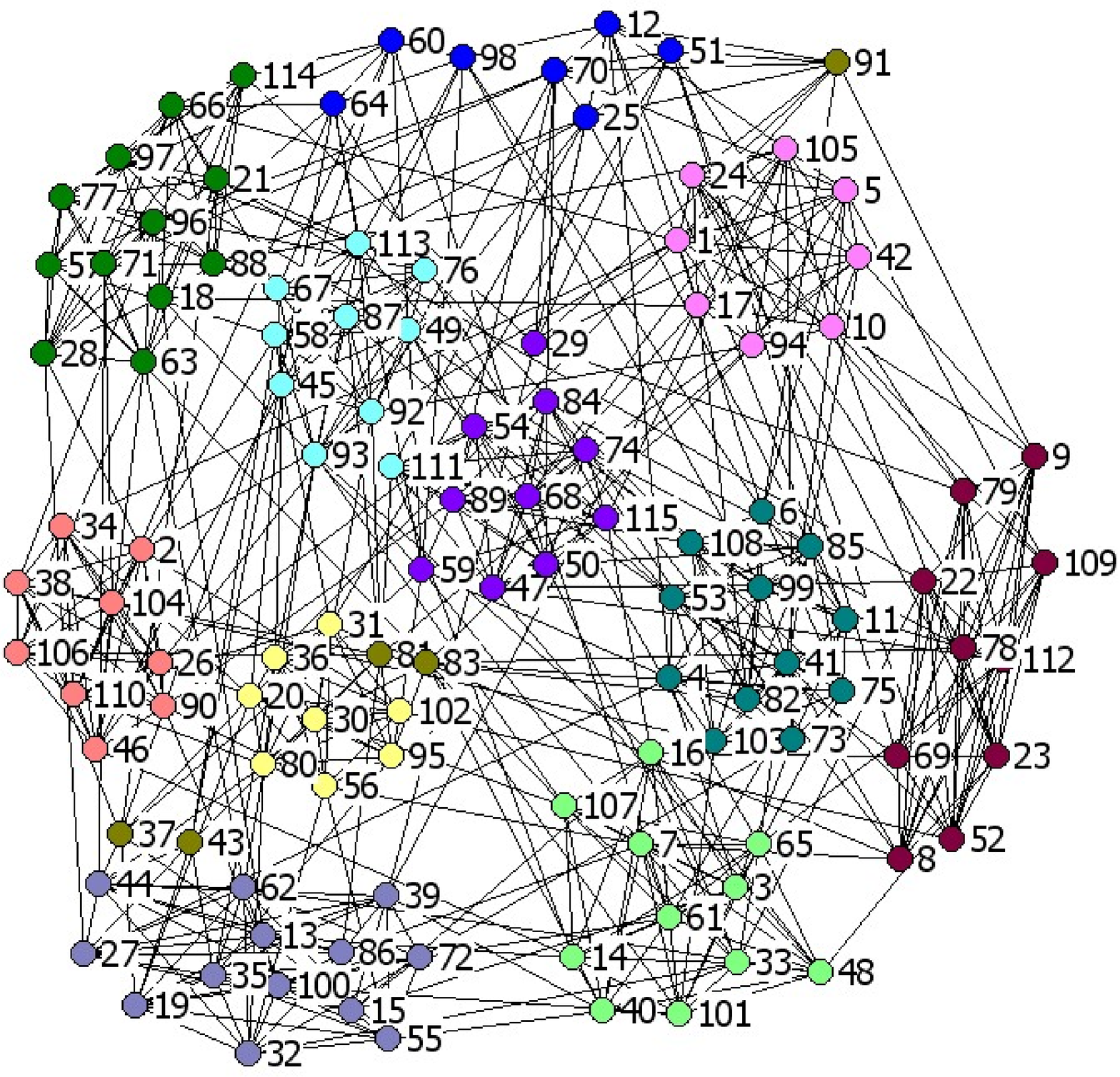}
\includegraphics[scale=0.30]{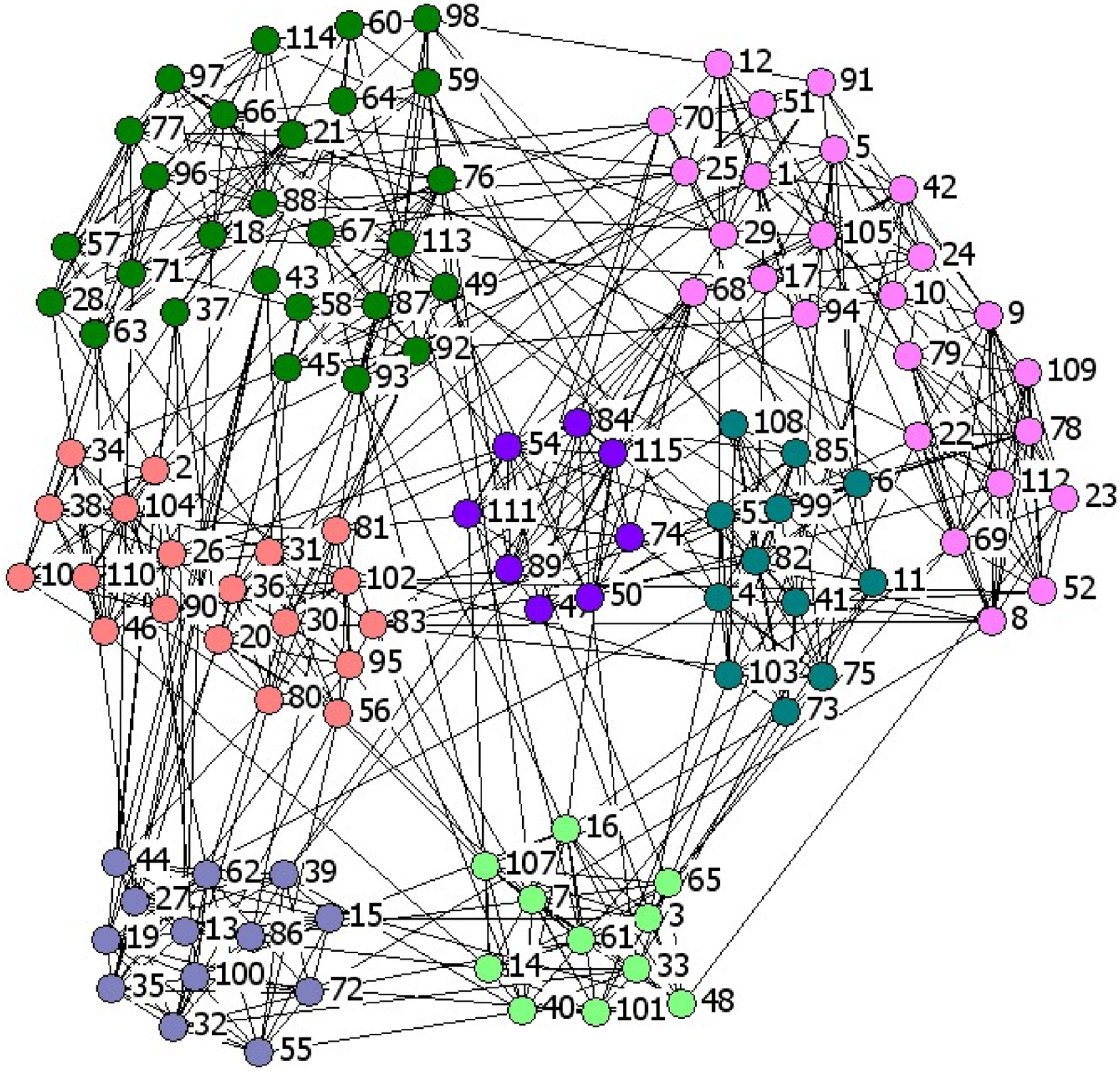}
\includegraphics[scale=0.30]{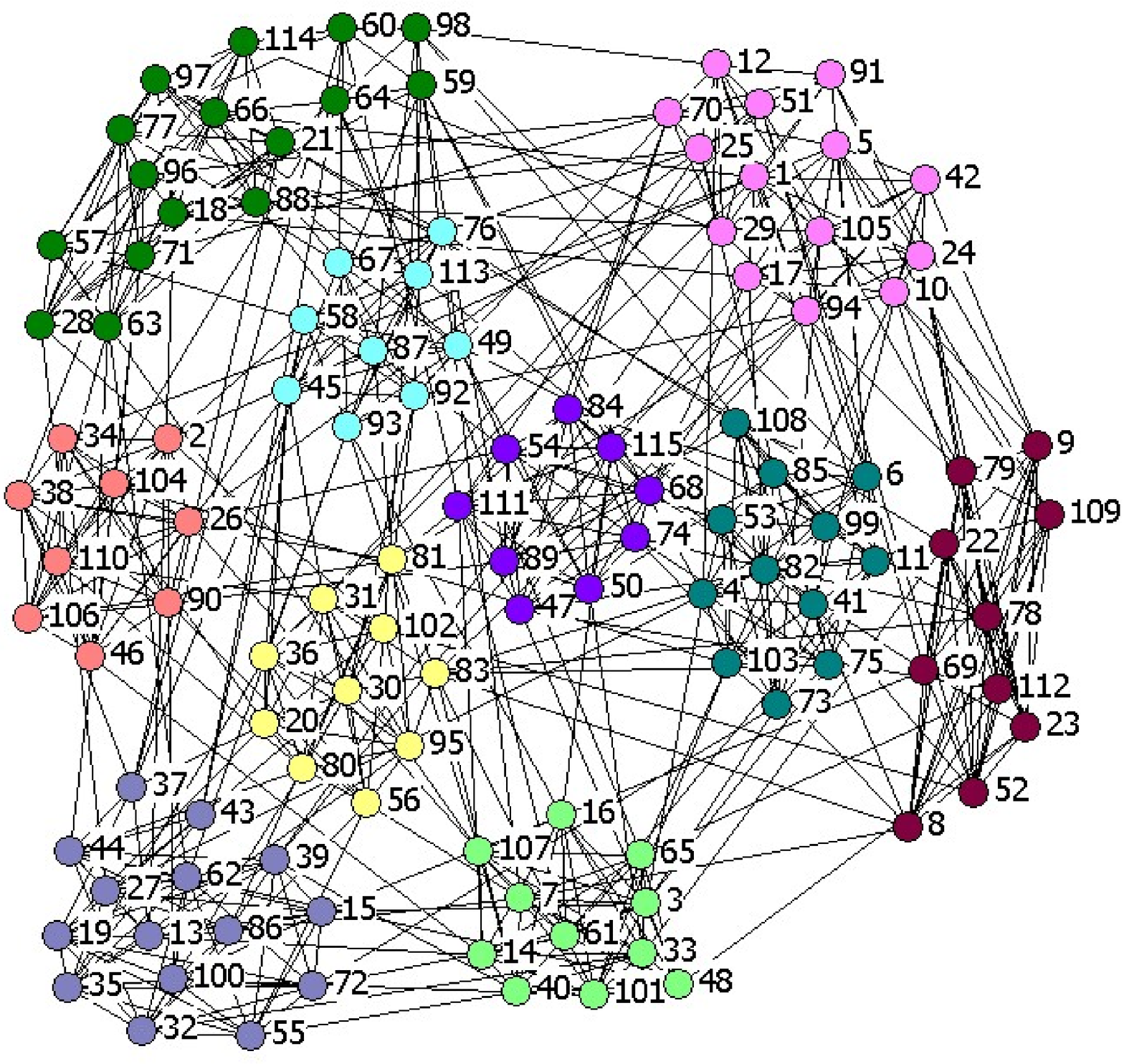}
\caption{Comparison of the algorithmic outputs corresponding to the
best identifications subject to modularity. The three panels are
(upper panel) real grouping in regular season Fall 2000, (middle
panel) resulting communities from the CNM algorithm, and (lower
panel) resulting communities from the XCZ+CNM algorithm. Each node
here denotes a football team and different colors represent
different groups/communities.}
\end{figure}
\end{center}

The CNM algorithm is relatively rough in the early stage, actually,
it strongly tends to merge lower-degree nodes together (see Eq. (2)
in Ref. \cite{Newman2004b}, the first term is not distinguishable in
the early stage while the enhancement of the second term favors
lower-degree nodes). This tendency usually makes mistakes in the
very early stage and can not be corrected afterwards. We therefore
propose a hybrid algorithm which starts from a $n$-division
$\Gamma_0=\{V_1,V_2,\cdots,V_n\}$, and takes the procedure mentioned
in the last paragraph for one round (i.e., step (i) and step (ii)).
The subgraph similarity is degenerated to the similarity between two
nodes:
\begin{equation}
s_{xy}=\frac{a_{xy}+n_{xy}}{\sqrt{k_xk_y}},
\end{equation}
where $n_{xy}$ denotes the number of common neighbors between $x$
and $y$, $a_{xy}$ is 1 if $x$ and $y$ are directly connected, and 0
otherwise. After this round, each subgraph has at least two nodes.
Then, we implement the CNM algorithm until all nodes are merged
together.

\section{Data}

In this paper, we consider five real networks drawn from disparate
fields: (i) Football.--- A network of American football games
between Division IA colleges during regular season Fall 2000, where
nodes denote football teams and edges represent regular season games
\cite{Girvan2002}. (ii) Yeast PPI.--- A protein-protein interaction
network where each node represents a protein
\cite{Mering2002,Bu2003}. (iii) Cond-Mat.--- A network of
coauthorships between scientists posting preprints on the
\emph{Condensed Matter E-Print Archive} from Jan 1995 to March 2005
\cite{Newman2001}. (iv) WWW.--- A sampling network of the World Wide
Web \cite{Albert1999}. (v) IMDB.--- Actor networks from the
\emph{Internet Movie Database} \cite{Ahmed2007}. We summarize the
basic information of these networks in Table 1.

\begin{table}
\caption{Basic information of the networks for testing.}
\begin{tabular}{p{2cm}p{3cm}p{3cm}p{2cm}}
\hline\noalign{\smallskip}
Networks & Number of Nodes, $|V|$ & Number of Edges, $|E|$ & References \\
\noalign{\smallskip}\svhline\noalign{\smallskip}
Football & 115 & 613 & \cite{Girvan2002}\\
Yeast PPI & 2631 & 7182 & \cite{Mering2002,Bu2003}\\
Cond-Mat & 40421 & 175693 & \cite{Newman2001} \\
WWW & 325729 & 1090107 & \cite{Albert1999}\\
IMDB & 1324748 & 3782463 & \cite{Ahmed2007} \\
\noalign{\smallskip}\hline\noalign{\smallskip}
\end{tabular}
\end{table}

\begin{table}
\caption{Maximal modularity.}
\begin{tabular}{p{2cm}p{1.7cm}p{1.7cm}p{1.7cm}p{1.7cm}p{1.7cm}}
\hline\noalign{\smallskip}
Algorithms & Football & Yeast PPI & Cond-Mat & WWW & IMDB \\
\noalign{\smallskip}\svhline\noalign{\smallskip}
CNM & 0.577 & 0.565 & 0.645 & 0.927 & N/A\\
XCZ & 0.538 & 0.566 & 0.682 & 0.882 & 0.691\\
XCZ+CNM & 0.605 & 0.590 & 0.716 & 0.932 & 0.786 \\
\noalign{\smallskip}\hline\noalign{\smallskip}
\end{tabular}
\end{table}

\begin{table}
\caption{CPU Time in millisecond (ms) resolution.}
\begin{tabular}{p{2cm}p{1.7cm}p{1.7cm}p{1.7cm}p{1.7cm}p{1.7cm}}
\hline\noalign{\smallskip}
Algorithms & Football & Yeast PPI & Cond-Mat & WWW & IMDB \\
\noalign{\smallskip}\svhline\noalign{\smallskip}
CNM & 172 & 5132 & 559781 & 12304152 & N/A \\
XCZ & 0 & 47 & 2022 & 17734 & 257875 \\
XCZ+CNM & 0 & 62 & 36422 & 443907 & 47714093 \\
\noalign{\smallskip}\hline\noalign{\smallskip}
\end{tabular}
\end{table}

\section{Result}

In Table 2 and Table 3, we respectively report the maximal
modularities and the CPU times corresponding to the CNM algorithm,
our proposed algorithm (referred as XCZ algorithm where XCZ is the
abbreviation of the authors' names), and the hybrid algorithm
(referred as XCZ+CNM). All computations were carried out in a
desktop computer with a single \emph{Inter CoreE2160} processor
(1.8GHz) and 2GB EMS memory. The programme code for the CNM
algorithm is directly downloaded from the personal homepage of
Clauset. The IMDB seems too large for the CNM algorithm, and we can
not get the result in reasonable time.

From Table 2, one can find that the XCZ algorithm can provide
competitively accurate division of communities verse the CNM
algorithm. A significant feature of the XCZ algorithm is that it is
very fast, in general more than 100 times fasters than the CNM
algorithm. Just by a desktop computer, one can find out the
community structure of a network containing $10^6$ nodes within
minutes. In comparison, the hybrid algorithm is remarkably more
accurate (measured by the maximal modularity) than both the CNM and
XCZ algorithms. In Figure 2, we compare the resulting community
structures of the Football network, from which one can see obviously
that the hybrid algorithm gives closer result to the real grouping
than the CNM algorithm. We think the hybrid algorithm is fast enough
for many real applications. Taking IMDB as an example, although it
contains more than $1.3\times 10^6$ nodes, the hybrid algorithm only
spends less than one day. Indeed, the hybrid algorithm outperforms
the CNM algorithm for both the accuracy and the speed.

\section{Conclusion}

Thanks to the quick development of computing power and database
technology, many very large scale networks, consisted of millions or
more nodes, are now available to scientific community. Analysis of
such networks asks for highly efficient algorithms, where the
problem of community identification has attracted more and more
attentions for its hardness and practical significance.

The agglomerative method based on node similarity \cite{Breiger1975}
is of lower accuracy compared with the divisive algorithms based on
edge-betweenness \cite{Girvan2002} and edge-clustering coefficient
\cite{Radicchi2004}. In this paper, we extended the similarity
measuring the structural equivalence of a pair of nodes to the
so-called \emph{subgraph similarity} that can quantify the proximity
of two subsets of nodes. Accordingly, we deigned an ultrafast
algorithm, which provides competitively accurate division of
communities while runs typically hundreds of times faster than the
well-known CNM algorithm. Using our algorithm, just by a desktop
computer, one can deal with a network of millions of nodes in
minutes. For example, it takes less than five minutes to get the
community structure of IMDB, which is consisted of more than
$1.3\times 10^6$ nodes.

Furthermore, we integrated the CNM algorithm and our proposed
algorithm and designed a hybrid method. Numerical results on
representative real networks showed that this hybrid algorithm is
remarkably more accurate than the CNM algorithm, and can manage a
network of about one million nodes in a few hours.

The modularity has been widely accepted as a standard metric for
evaluating the community identification, as well as has found some
other applications such as being an assistant for extracting the
hierarchical organization of complex systems \cite{Sales-Pardo2007}.
Although modularity is indeed the most popular metric for community
identification, and the result corresponding to the maximal
modularity looks very reasonable (see, for example, Figure 2), it
has an intrinsic resolution limit that makes small communities hard
to detect \cite{Fortunato2007,Lancichinetti2008}. An alternative,
named \emph{normalized mutual information} \cite{Danon2005} is a
good candidate for future investigation. In addition, an extension
of modularity for weighted networks, namely \emph{weighted
modularity} \cite{Newman2004c}, has been adopted to deal with
community identification problem in weighted networks
\cite{Fan2007,Mitrovic2008}. We hope the subgraph similarity
proposed in this paper can also be properly extended to a weighted
version to help extract the weighted communities.

\begin{acknowledgement}
This work is benefited from the \emph{Pajek Datasets} and the
\emph{Internet Movie Database}, as well as the network data
collected by Mark Newman, Albert-L\'aszl\'o Barab\'asi and their
colleagues. E.-H.C. acknowledges the National Natural Science
Foundation of China under grant numbers 60573077 and 60775037. T.Z.
acknowledges the National Natural Science Foundation of China under
grant number 10635040.
\end{acknowledgement}

\end{document}